\documentclass[aps,prl,twocolumn,superscriptaddress,
nobibnotes,nodoi,noeprint,longbibliography]{revtex4-1}
\usepackage{graphicx}
\usepackage{dcolumn}
\usepackage{bm}
\usepackage{float}
\usepackage{color}
\usepackage{refstyle}
\usepackage{mathrsfs}
\usepackage{amsmath}
\usepackage{esint}
\usepackage{siunitx}
\usepackage[unicode=true,pdfusetitle,bookmarks=true,
bookmarksnumbered=false,bookmarksopen=false,breaklinks=false,
pdfborder={0 0 0},backref=false,colorlinks=true,citecolor=red]
{hyperref}
\DeclareMathOperator{\sgn}{sgn}

\begin{document}
\title{Bidirectional Zigzag Growth from Clusters of Active Colloidal Shakers}
\author{Gaspard Junot}
\affiliation{Departament de F\'{i}sica de la Mat\`{e}ria Condensada, Universitat de Barcelona, 08028 Spain}
\author{Andr\'{e}s Javier Manzano Gonz\'{a}lez}
\affiliation{Departament de F\'{i}sica de la Mat\`{e}ria Condensada, Universitat de Barcelona, 08028 Spain}
\author{Pietro Tierno}
\email{ptierno@ub.edu}
\affiliation{Departament de F\'{i}sica de la Mat\`{e}ria Condensada, Universitat de Barcelona, 08028 Spain}
\affiliation{Universitat de Barcelona Institute of Complex Systems (UBICS), Universitat de Barcelona, Barcelona, Spain}
\affiliation{Institut de Nanoci\`{e}ncia i Nanotecnologia, Universitat de Barcelona, Barcelona, Spain}
\date{\today}
\begin{abstract}
Driven or self-propelling particles moving in viscoelastic fluids recently emerge as novel class of active systems showing a complex yet rich set of phenomena due to the non-Newtonian nature of the dispersing medium. Here we investigate the one-dimensional growth of clusters made of active colloidal shakers, which are realized by oscillating magnetic rotors dispersed within a viscoelastic fluid and at different concentration of the dissolved polymer. These magnetic particles when actuated by an oscillating  field display 
a flow profile similar to that of a shaker force dipole, i.e. without any net propulsion. We design a protocol to assemble clusters of colloidal shakers and induce their controlled expansion into elongated zigzag structures. 
We observe a power law growth of the mean chain length
and use theoretical arguments to explain the measured $1/3$ exponent. These arguments agree well with both experiments and particle based numerical simulations. 
\end{abstract}
\maketitle
\subsection*{Introduction}
Investigating the formation of dynamic patterns 
from a collection of active or self-propelling particles 
is a rich research field that has led to the 
observation of fascinating phenomena
including swarming~\cite{Vicsek1995,Yang2010,Yingzi2010,Zion2022}, clustering~\cite{Peruani2006,Wensink2008,Peruani2011,Pohl2014,Ginot2015,Shoham2023}, crystallization~\cite{Julian2012,Weber2014},
dynamic vortices and swirls~\cite{Kudrolli2008,Kaiser2017,Nishiguchi2018} or  
phase-separation induced by motility~\cite{Redner2013,Buttinoni2013,Linden2019,Caballero2022} among others. 
Moreover, collective ensemble of active particles that can be controlled by an external field may be used as "progammable matter" 
to preform useful tasks at the microscale,   
with potential applications in robotics~\cite{Gross2006,Rubestein2014,Miskin2020}, microfluidics~\cite{Snezhko2011,Sanchez2011} 
or material science~\cite{Simmchen2022}. 

While most of the prototypes realized so far have been 
dispersed in Newtonian fluids, such as water, 
many new interesting effects may arise when the fluid medium is non-Newtonian, such as a viscoelastic one~\cite{Henry2007,Lauga2007,Teran2010,Gomez2016,Narinder2018,Kai2020,Spagnolie2023}. 
Indeed,
in biological systems microorganisms such as sperm cells 
navigate in a non-Newtonian fluid. 
The non-linearity of the dispersing medium may affect 
the sperm transport~\cite{Tung2017} apart from being important 
in several other processes including biofilm formation~\cite{Flemming2010,Houry2012} or fertilization~\cite{Bigelow2004}. 
As previously reported, 
a viscoelastic medium may even induce propulsion to 
a reciprocal swimmer
which performs periodic, time-reversible, body-shape deformations~\cite{Qiu2014}.

In a recent experimental work~\cite{junot2023large}, we reported the formation of large scale zigzag bands made of a population of magnetic rotors which were reversibly actuated by an external, oscillating magnetic field. When the magnetic rotors oscillate in water, the  particles perform periodic back-forward rolling being unable to organize in any significant structure, i.e. they remain evenly distributed across the plane. In contrast, by adding few amount of polymer that makes the medium viscoelastic, we observe that the particles self-organize into zigzag structures, that merge in time
perpendicular to the direction of the oscillating field. 
The progressive coarsening of these bands would ultimately lead to the formation of a single chain of particles with the size of the system. 
In our previous work~\cite{junot2023large}, it was not possible to investigate the elongation dynamics and reach the steady state since thick bands, as the one shown in Figure 1(b), extend above the observation area. Moreover, these bands reached the boundaries of the experimental cell, thus interacting with the confining walls. To study the evolution of the system toward its steady state, we have developed a protocol to create isolated cluster from which smaller bands growth, and sufficiently far from neighboring bands and from the confining walls. Under these conditions, we report a growth process that could not be observed in Ref.~\cite{junot2023large}. Indeed, when isolated, thin bands laterally extend while reducing their thickness over time.
We observe a power law growth of these lines and analyze in details the influence of the polymer concentration on the velocity field generated by a particle as well as on the growth process. Finally, using a simple theoretical argument based on conservation law, we explain the growth and the observed $1/3$ exponent. We confirm these predictions by doing particle based  numerical simulations that agree both with the model and the experiments.

\begin{figure*}[t]
\begin{center}
\includegraphics[width=\textwidth,keepaspectratio]{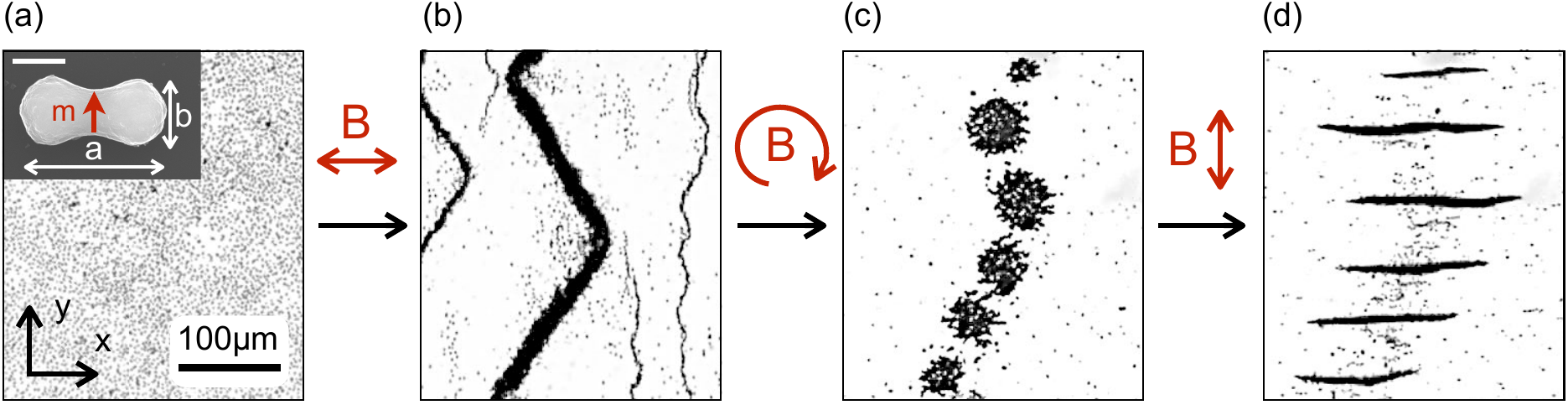}
\caption{(a-d) Sequence of experimental images showing the protocol employed to generate growing clusters of colloidal shakers. We start from a homogeneous dispersion of particles in a mixture of water  and PAAM,  as shown in (a). We first apply a time dependent rotating field along the $\bm{x}$ direction,  Eq.~\ref{applied_field}, which induces the grow of zigzag bands along the perpendicular one ($\bm{y}$), as shown in (b).
After that, an in plane rotating magnetic field is used to break the band
into a sequence of circular clusters (c). Then the oscillating field 
is switched along the $\bm{y}$ direction, and it induces the grow 
of parallel zigzag structures, as shown in (d). Details on the field parameters are given in text, the scale bar for all images is $100 \, {\rm \mu m}$. Top inset in (a) shows a scanning electron microscope image of a 
hematite particle with dimensions ($a= 2.6 \, {\rm \mu m}$, $b=1.2\,  {\rm \mu m}$) and the direction of permanent moment $m$.
The scale bar is $1 \, \rm{\mu m}$. The sequence of images (c,d) is shown in VideoS1 in~\cite{EPAPS}.}
\label{figure1}
\end{center}
\end{figure*}

\subsection*{Experimental system}
Our colloidal shakers are realized 
by 
cyclically actuating anisotropic hematite microparticles 
in a viscoelastic medium. As shown in the 
top inset of Fig.~\ref{figure1}(a), 
the hematite are characterized by 
two connected spherical lobes of equal diameters $b=1.2\,  {\rm \mu m}$
with a total length 
of $a=2.6 \, {\rm \mu m}$.
This peanuts-like shape of the particles is the result of their chemical synthesis, performed following the sol gel approach~\cite{Sugimoto1993},
more detail can be found in a previous work~\cite{Martinez16}.
The particles are ferromagnetic and characterized by a permanent moment 
directed perpendicular to their long axis,
with magnitude  $m \simeq 9 \cdot 10^{-16} \, \rm{Am^2}$~\cite{Martinez-Pedrero034002}.
We disperse these particles in
a viscoelastic medium made of a water solution 
of polyacrylamide (PAAM), a water soluble high-molecular weight polymer
($M_w = 5-6 \cdot 10^6$).
In this work we change the  polymer concentration
within the range  $c_p \in [0, 0.05]\%$ in weight, relative to water.
For such concentration values, the PAAM solution can be considered 
in the dilute
regime, where the polymer chains do not overlap. Previous works estimate 
the transition between dilute and semidilute regime, e.g. the
polymer chains overlap without entanglement, at 
$c \sim [0.06-0.1]\%$ \cite{zell2010there,del2015rheometry}. 
This dilute regime was chosen for the relative low viscosity,
similar to that of water, that allow to easily manipulate the magnetic particles.

We mix the particles and the PAAM solution, 
the resulting suspension is then confined between 
a plastic petridish and a cover slip that are later sealed. The
experimental cell, with a final thickness of $\sim 260 \rm{\mu m}$ is placed on the stage of a custom made optical microscope.
The latter is connected to a charge-coupled device (ccd) camera (Scout scA640-74f, Basler) that allow to record real-time videos of the particle dynamics at $75$ frame per seconds.
Videos of the growth process have been taken at a lower frame rate of $30$Hz. 
Further, the microscope is equipped with 
a set of custom-made magnetic coils having their axis
aligned along the three orthogonal directions $(\bm{\hat{x}},\bm{\hat{y}},\bm{\hat{z}})$. 
We generate a rotating magnetic field 
perpendicular to the
substrate plane ($\bm{\hat{x}},\bm{\hat{z}}$) by connecting three of coils
to a power amplifier (IMG STA-800, Stage Line)  that was controlled via a arbitrary waveform generator (TGA1244,
TTi).

\begin{figure*}[th]
\begin{center}
\includegraphics[width=\textwidth,keepaspectratio]{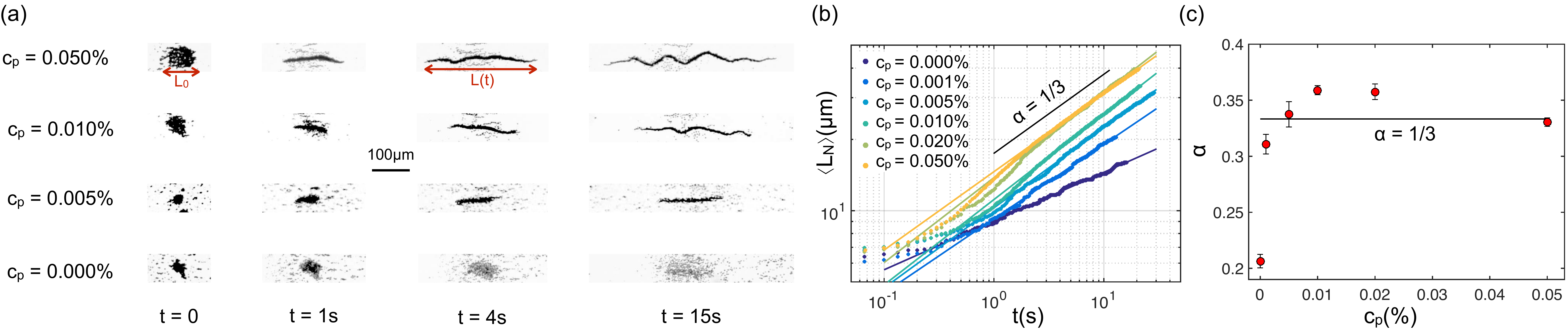}
\caption{(a) Table illustrating the evolution of 
clusters with initial length $L_0$ into bands of length $L(t)$ 
for different time $t$ and amount of dispersed polymer (PAAM) given by $c_p$. As shown in the first column, the used concentration $c_p\in [0,0.05]\%$. The oscillating field (Eq.~\ref{applied_field}) is applied at $t=0$, and 
is the same for all PAAM concentrations. 
See corresponding VideoS2 in~\cite{EPAPS}
for the case in pure water, showing the absence of zigzag structures. (b) Mean normalized length $\langle L_N \rangle= \langle L/N^{1/3}\rangle$ as function of time $t$ in log-log scale  for different PAAM concentrations. Dots are experimental data, the solid lines are linear regression used to extract the exponent $\alpha$. (c) Exponent $\alpha$ as a function of $c_p$.}
\label{figure2}
\end{center}
\end{figure*}

\subsection*{Realization of parallel zigzag bands}
Once dispersed in the polymer solution, the hematite particles sediment close to the bottom plane, and display weak thermal fluctuations.
To realize the colloidal shakers, we drive these particles back and forward along a fixed direction (here the $\bm{\hat{x}}$-axis) using a time dependent rotating field,
\begin{equation}  
\bm{B} = B\left[\sin{(2\pi t \Delta f_{-}})\bm{\hat{x}} + \cos{ (2\pi t\Delta  f_{+})}\bm{\hat{z}}\right] \, \, \, .
\label{applied_field}  
\end{equation}  
Here $B$ is the field amplitude, which we fix to $B=5.5$mT, $f$ the driving frequency also fixed to $f=80$Hz,
$\Delta f_{\pm}= f \pm \delta f/2$
and $\delta f$ the frequency difference between the two fields components along the $\bm{\hat{x}}$ and $\bm{\hat{z}}$ axis, $\delta f = 4$Hz.
The field in Eq.~\ref{applied_field} periodically changes the direction of rotation each  
$\delta t = 1/ (2\delta f)$, and this effect has two consequences.
First it aligns the permanent moments of the particles imposing a fixed orientation with respect to their long axis.
Second, the particles are subjected to a magnetic torque
$\bm{\tau}_m=\bm{m} \times \bm{B}$
which set them in rotational motion along their 
short axis at an angular speed $ \Omega  = 2\pi f$ and close to the bottom plane.
Due to the rotational-translation hydrodynamic coupling~\cite{Happel1973}, the spinning hematite
performs periodic displacements back and forward
following synchronously the field rotations.
Indeed, we have independently check that, for driving frequencies $f<100$Hz,  the particles are  in the synchronous regime, i.e. their magnetic moment is locked to the field by a constant phase-lag angle~\cite{Tierno2009,Junot2021}.

As shown in the sequence of images in Figure~\ref{figure1}(a,b),
when the particles are homogeneously dispersed in a PAAM solution at a concentration $c_p=0.05\%$, the applied modulation along the $\bm{\hat{x}}$ direction induces the growth
of large scale zigzag bands along the perpendicular, $\bm{\hat{y}}$
direction. As time proceeds, 
the bands 
coarsen by acquiring neighboring particles and merge with nearest bands, forming one large structure that extends beyond the microscope observation area. Depending on the initial particle concentration, this structure can reach 
a length of few $mm$ and a thickness of $\sim 50 \rm{\mu m}$
Figure~\ref{figure1}(b).  
Our protocol to create localized bands, such that the lateral growth (along $\bm{\hat{x}}$) process can be 
entirely visualized, consists in 
transforming a large zigzag structure into a series of small clusters 
via application of an in-plane rotating magnetic field, 
$\bm{B}\equiv B [\cos{(2 \pi f t)} \bm{\hat{x}}+\sin{(2 \pi f t)} \bm{\hat{y}}]$, Fig.~\ref{figure1}(c).
As previously reported for different magnetic colloids in water~\cite{Tierno2007,Osterman2009} or non-magnetic particles in a ferrofluid~\cite{Pieranski1996},
the rotating field applies a torque to each particle and also induces time-average dipolar attractions, which are not affected by the presence of the PAAM. 
Thus, the large band breaks into several rotating circular clusters 
composed of attractive spinning colloids.
The size of these clusters result from the balance between 
the magnetic attraction and the repulsive hydrodynamic flow 
induced by the particle spinning motion~\cite{massana2021arrested}.
After that, we apply again the oscillating field,  Eq.~\ref{applied_field},
but now along the perpendicular $\bm{\hat{y}}$ direction, inducing the formation of parallel lines of shakers growing along the $\bm{\hat{y}}$ axis, 
Figure~\ref{figure1}(d), see also VideoS1 in~\cite{EPAPS}.

The table in Fig.~\ref{figure2}(a) illustrates how the PAAM concentration affects the band growth at different times. 
For the largest amount of polymer tested ($c_p=0.05\%$),
we find that a circular cluster first collapses into a line in less than a second, the line then lengthens and  becomes thinner 
since the number of particles is constant. After $\sim 5$s, the line deforms and it acquires a zigzag shape  where branches are arranged at a constant angle of $\theta_l=\pm 31^{\circ}$. This effect was explained in Ref~\cite{junot2023large} 
by considering the shaker-like shape of the flow field generated by the particle rotation. Decreasing the PAAM concentration reduces the angle of the zigzag structures. While at high PAAM concentration the particles self-organize into zigzags with sharp tips merging branches at a constant angle, reducing the PAAM concentration the stripes flatten, reducing this angle. Note that this effect occurs for all initial particles configurations, whether the particles are homogeneously distributed across the plane as shown in Fig.~\ref{figure1}(a), or when they form localized clusters, as shown in Fig.~\ref{figure2}(c). In pure water
the cluster of oscillating particles does not form any band but rather grow uniformly  due to diffusion, bottom row in  Fig.~\ref{figure2}(a),
and also illustrated in VideoS2 in~\cite{EPAPS}. 

\subsection*{Band elongation}
To characterize the longitudinal growth of the clusters, we measure the length $L(t)$ as function of time $t$ for different polymer concentrations $c_p$. In  Fig.~\ref{figure2}(b)
we divide $L$ by $N^{1/3}$ as, $\langle L_N \rangle  = \langle L/N^{1/3} \rangle$ being
$N$ the number of particle within a cluster. We use this rescaling since the initial condition, i.e. the initial number of particles $N$, is difficult to control experimentally. To compute $N$, we measure the initial cluster diameter $L_0$ and $N = (\pi (L_0/2)^2)/a_p$, being $a_p$ the area covered by a single particle.  
All curves shown in Fig.\ref{figure2}(b) display two distinct regimes: first $\langle L_N \rangle $ barely increases for times shorter than \SI{1}{\second}, which corresponds to the initial collapse of the cluster in a line. After that, the line grows as 
a power law with an exponent $\alpha= 1/3$ for all $c_p$ except for pure water, where the dynamics are slower as governed by diffusion and possible weak hydrodynamic interactions, and  $\alpha = 0.210 \pm 0.006$, Fig.\ref{figure2}(c).  
We note that, the exponent observed in our system is in general smaller than that measured when magnetic colloids form chains in water and under a static field, which was $\alpha \geq 0.5$~\cite{Faraudo2016}. This indicates a different growth mechanism which, in our case,  is due to the presence of other interactions than magnetic dipolar ones.

\begin{figure}[th]
\begin{center}
\includegraphics[width=\columnwidth,keepaspectratio]{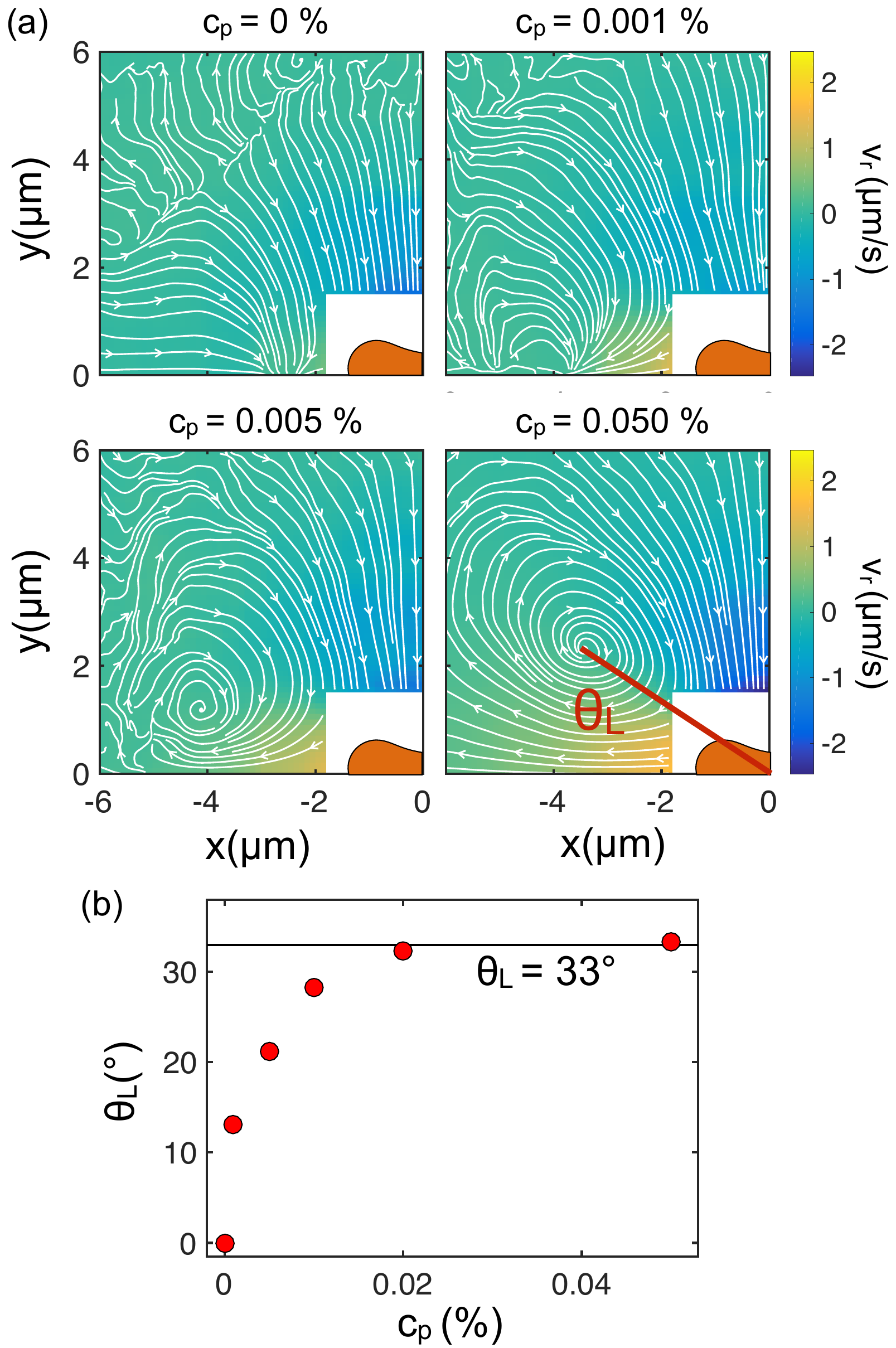}
\caption{(a) Panels showing the flow fields generated by a shaker for different PAAM concentrations. We plot the radial velocity 
$v_r$ in the $(\bm{\hat{x}},\bm{\hat{y}})$ plane of one shaker with the streamlines superimposed for four different PAAM concentration. Without PAAM, the flow field is mainly attractive and the fluid flows towards the microrotor. As one raises $c_p$, a repulsive zone starts to appear leading to the formation a vortex. The attractive and repulsive zones are separated by a line making an angle $\theta_l$ with the horizontal axis. (b) Limit angle $\theta_L$ as function of $c_p$.}
\label{figure3}
\end{center}
\end{figure}

To get more insight on the elongation behavior, 
we use particle tracking velocimetry
to measure the flow field profile generated by the rotation of a single shaker as a function of $c_p$, 
Fig.\ref{figure3}(a), we provide more details in Appendix A. Without PAAM, the flow field is weakly attractive everywhere except at short distances, i.e. close to the particle tip where the streamlines converge to a stagnation point near the tip. Indeed, in a previous work~\cite{Martinez2018} 
it was shown that close propelling hematite particles can create a hydrodynamic bound state where they align tip to tip,
i.e. in agreement with the presence of this stagnation point. 
By adding PAAM the situation changes 
since elastic effects due to the polymer start to appear. In this situation, the flow profile become similar to that of 
a shaker-like force dipole~\cite{hatwalne2004rheology},
which is attractive at the particle side and repulsive along the two tips. 
It was shown previously using a Oldroyd-B constitutive model~\cite{junot2023large}, that this flow field is the consequence of the first normal stress induced by the particle rotating in the viscoelastic fluid.
By increasing the PAAM concentration, we observe that a vortex starts forming close to the particle tips ($\theta \approx 0$) and progressively moves to finally reach a position at $\theta \simeq \SI{33}{\degree}$. As shown in Fig.~\ref{figure3}(b), an attractive and repulsive zone are
separated by an angle $\theta_L \in [0,33]^{\circ}$
for $c_p \in [0,0.05] \%$.
As $\theta_L$ decreases, the zigzag flatten into a line, Fig.~\ref{figure2}(a), $c_p = 0.005\%$, and it disappear for $c_p = 0\%$, Fig.~\ref{figure2}(a), bottom row.

\begin{figure}[th]
\begin{center}
\includegraphics[width=\columnwidth,keepaspectratio]{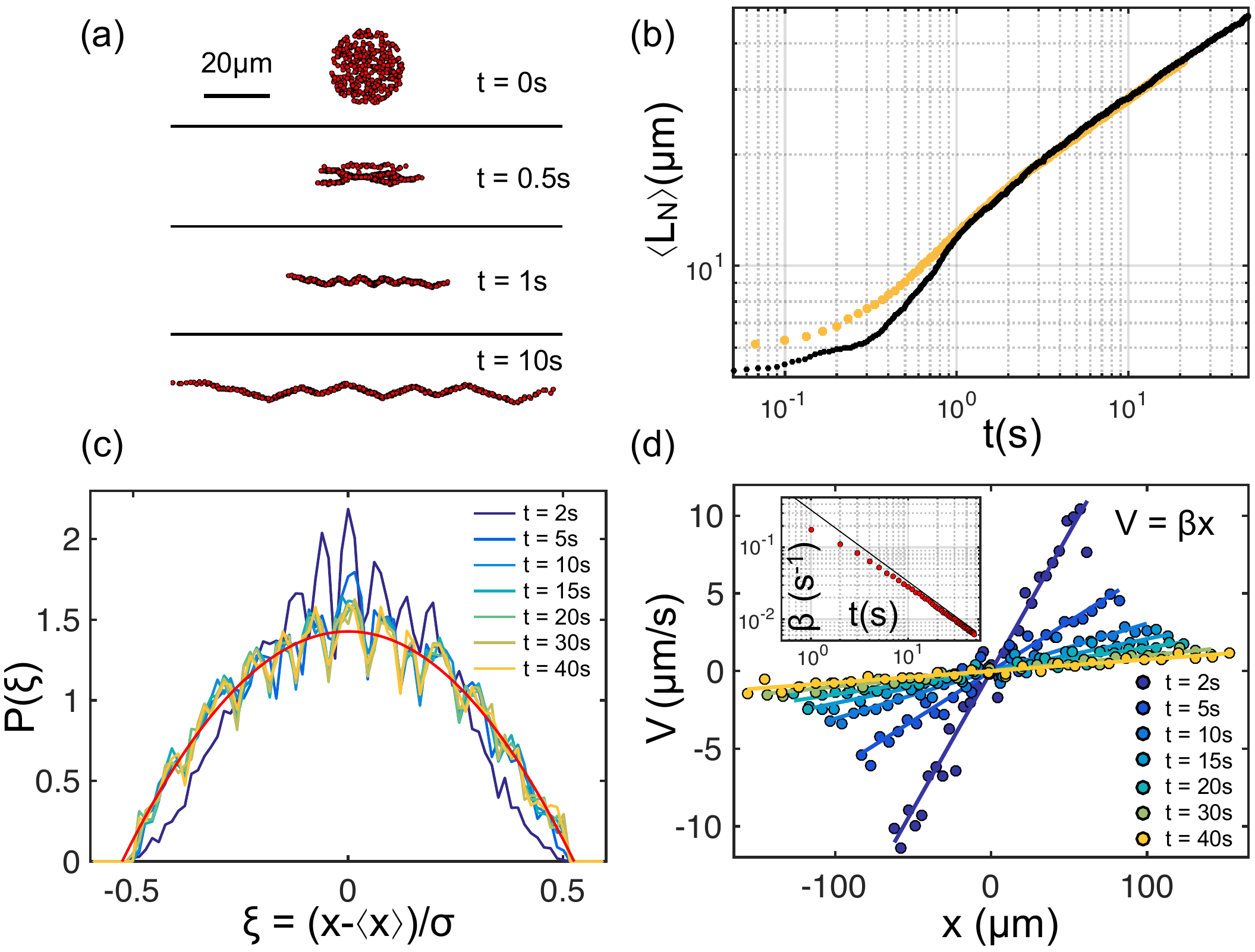}
\caption{Snapshots of a simulation at different times $t$. (b) Normalized mean length $\langle L_N \rangle$ as function of time, the experimental curve for $c_p=0.05\%$ is in orange, the back curve is the results of the simulation. (c) Normalized distribution of particle $P(\xi)$ at different times $t$. The red curve is a parabolic fit of the curve at $t=\SI{40}{\second}$. (d) Local velocity $v_x$ as function of $x$ and at different times $t$. Points are results of the simulation, the solid lines are linear fit used to extract the slope $\beta$ as function of time. Inset: slope $\beta$ as function of time, dots are results of the simulation, the black line represents the function $y(t) = 1/3t$. }
\label{figure4}
\end{center}
\end{figure}

\subsection*{Theory}
We consider the 
zigzag band as a one dimensional ($1$D) chain of particles characterized by a spatial density $\rho(x,t)$ and a local velocity $v(x,t)$. This velocity results from the hydrodynamic flow fields generated by all nearest particles. Moreover, since we reduce the problem to $1$D, the interactions between the particles are purely repulsive ($\theta = 0$).
Thus, the local velocity at position $x$ is the integral of all the particles velocity:
\begin{equation}
v (x,t)= \int_D v_p(x-x')\rho(x',t)dx' 
\end{equation}  
being $v_p(x-x')$ the flow velocity that a particle at position $x$ exerts on a particle at position $x'$ and $D$ is the domain of integration.
Further, to simplify the problem,we assume that the flow velocity generated by  a particle at a position $x$ is equal to a constant $v_p$ within the interval $[x-r_c~ ; ~ x+r_c]$ and zero elsewhere, being $r_c$ a cutoff length. Thus, we have:
\begin{equation}
v (x,t)= v_p \int_{x-r_c}^{x+r_c} \sgn(x-x') \rho(x',t)dx' 
\label{eq2}
\end{equation}  
being $\sgn$ is the signum function.
Since the zigzag band can be considered as isolated due to the relative large distance with nearest bands, as shown in Fig.~\ref{figure1}(d), the total mass within a band is conserved:
\begin{equation}
\frac{\partial \rho}{\partial t} = - \frac{\partial(v\rho)}{\partial x}
\label{eq3}
\end{equation} 
and we search for a solution $(\rho(x,t),v(x,t))$ that satisfies both Eqs.\ref{eq2},\ref{eq3}. 
A solution of this problem is given by:
\begin{equation}
\begin{aligned}
\label{eq4}
\rho(x,t) &= \frac{-x^2}{6 v_p r_c^2 t} + \left(\frac{3N^2}{32 r_c v_p t}\right)^{1/3}\\
v(x,t) &= \frac{x}{3t} \, \, \, ,
\end{aligned}
\end{equation} 
more details are given in Appendix B.
Since there are no particles outside the chain, the density 
$\rho(x,t)$ 
is zero at $x=\pm L/2$. Imposing this boundary condition, we arrive at:
\begin{equation}
\label{eq5}
L(t) = \left(36 N v_p r_c^2 t \right)^{1/3}
\end{equation} 
and $L_N \propto t^{1/3}$.
Eq.~\ref{eq4} shows that the velocity scales linearly with space with a slope $1/3t$, and the density $\rho$ is a parabola which becomes flat and wider with time.
 
To confirm these results, we have set-up a minimal simulation scheme that reproduces the zigzag band growth 
using the velocity field from the experimental data.
In particular,  we obtained the relative velocity $\bm{v}_{exp}(r_{ji},\theta_{ji})$ between two particles as a function of their relative distance $r_{ji}$ and angle $\theta_{ji}$. These measurements were performed in a previous study~\cite{junot2023large}.
Then, we fit the experimental data with an empirical function (see Appendix C for more details), and integrate the corresponding equation of motion:
\begin{equation}\label{numeq1}
\frac{d\bm{r}_i}{dt} = \sum_{j\neq i} \bm{v}(r_{ji},\theta_{ji}) 
\end{equation}
being $\bm{v}$ the velocity field around a particle.
As initial condition, we consider $N$ particles randomly distributed in a disk of diameter $L_0=\SI{40}{\micro\meter}$ and at a packing fraction of $\phi = 0.7$, and we use as cutoff length $r_l=\SI{20}{\micro\meter}$ beyond which the particles no longer interact, 
see Appendix C for more detail about the simulations. Note that when we set  $r_l<L_0/2$, sometimes we observe in our simulations that the cluster collapses into two distinct bands which merge at a later stage. For $r_l \geq L_0/2$ instead,  the cluster always collapses into a single band.

The simulations reproduce quantitatively the experimental results, as shown in Figs.~\ref{figure4}(a-b).
Indeed, the sequence of images in Fig.~\ref{figure4}a 
display the behavior of an ensemble of shakers 
where the initial cluster collapses first into a line within $\sim 1$ second and later it lengthens with time similar to the experimental observations.
For $c_p = 0.05\%$ the length $L_N$ matches very well the experimental data showing the scaling $t^{1/3}$,
as shown in Fig.~\ref{figure4}(b).
We further use the simulations to test the 
predictions of the model for $\rho(x,t)$ and  $v(x,t)$. 
Indeed, compared to the experiments, the simulations allow to resolve all particle positions and speeds, a task that is difficult experimentally  due to the large system density. 
To confirm the predicted behavior of $\rho(x,t)$, we have computed the distribution of the normalized position $\xi = (x-\langle x \rangle)/\sigma$ at different times $t$, being $\sigma$ the standard deviation of the particle positions. After a short transient regime, all distributions tend to collapse into a parabola, as predicted by the model, Fig.~\ref{figure4}(c). Also we observe that the local velocity $v(x,t)$ scales linearly with the position $x$ displaying a slope equal to $1/3t$, as shown in the inset of Fig.~\ref{figure4}(d). 

\subsection*{Conclusions}
We have investigated the growth process of clusters made of shaking 
magnetic rollers which elongate into a zigzag structure within a viscoelastic medium. We observe that circular clusters first collapse into a line and later grow with time following a power-law with an exponent $1/3$.  The collapse find its origin in the attractive part of the flow field, Fig.~\ref{figure3}, as reveled by particle tracking velocimetry. Thus, particles located at a relative angle $\theta > \SI{33}{\degree}$ attract each others and end up side by side while the interaction is repulsive for $\theta < \SI{33}{\degree}$. We simplify the problem by considering a one dimensional model and find that a constant repulsion 
between the shakers with a cutoff length is sufficient to explain the power-law growth and corresponding $1/3$ exponent.
These predictions were tested numerically finding a good agreement. 
Moreover, the growth scenario remain similar by changing the added concentration of polymer, except for the case $c_p = 0\%$.

The possibility of controlling the linear growth of dense clusters of active particles may find different technological applications. For example, zigzag bands display a conveyor belt current along their edges able to drag non-magnetic particles. Such localized hydrodynamic flow could be used to transport chemicals or biological species along a desired location in a microfluidid chip. 
The use of rotating or translating magnetic inclusions to generate controlled flows within a microfluidic environment was demonstrated 
in previous works in Newtonian media~\cite{Terray2002,Sawetzki2008,Tierno2008,Kavcic2009,Kavre2014,Pedrero2015}. We here show that these effects can be further extended to viscoelastic fluids, thus opening the doors towards direct application in biological systems.    
 
\subsection*{Acknowledgments}
We thank Marco De Corato for many stimulating discussions on the subject of this work.
This project has received funding from the 
European Research Council (ERC) under the European Union's Horizon 2020 research and innovation program (grant agreement no. 811234).  P.T. acknowledges support from the 
Ministerio de Ciencia e Innovaci\'o
(Project No. PID2022-137713NB-C22), the
Ag\`encia de Gesti\'o d'Ajuts Universitaris i de Recerca (Project No. 2021 SGR 00450)
and the Generalitat de Catalunya under Program ``ICREA Acad\`emia''.

\bibliography{Bibliography}

\section{Appendix A: particle tracking velocimetry}
We obtain the particle flow fields shown in the panels of Fig.~\ref{figure3}(a) by performing particle tracking 
of passive, tracer spheres, dispersed with the peanut particles.
The flow fields  were obtained from a dilute solution of peanut particles, so that interactions between them are negligible, mixed with silica colloids of $\SI{1}{\micro\meter}$ diameter that were used as tracers. 
We then record $26$ videos of $1$min duration at $75$fps using an 
oil immersion $100\times$ Nikon objective.
For these videos  we tracked the position of both the peanut and the tracers. In particular, we considered a square region of $\SI{10}{\micro\meter}$ around the peanut. This region was divided in square cells of lateral size $d=\SI{0.05}{\micro\meter}$. Inside these cells, the local mean relative velocities (velocity along $\hat{\bm{x}}$ and $\hat{\bm{y}}$ and radial $\hat{\bm{r}}$ direction) between the peanut and the tracers are computed by averaging over all the particles inside the cell, over time and the different experimental videos). We then performed a spatial moving average with a square windows of size $21d = \SI{1.05}{\micro\meter}$.
Since the flow field is symmetric, we further averaged the four quadrant of the image.

\section{Appendix B: solution of Eq.4 }

\begin{figure}[th]
\begin{center}
\includegraphics[width=7cm,keepaspectratio]{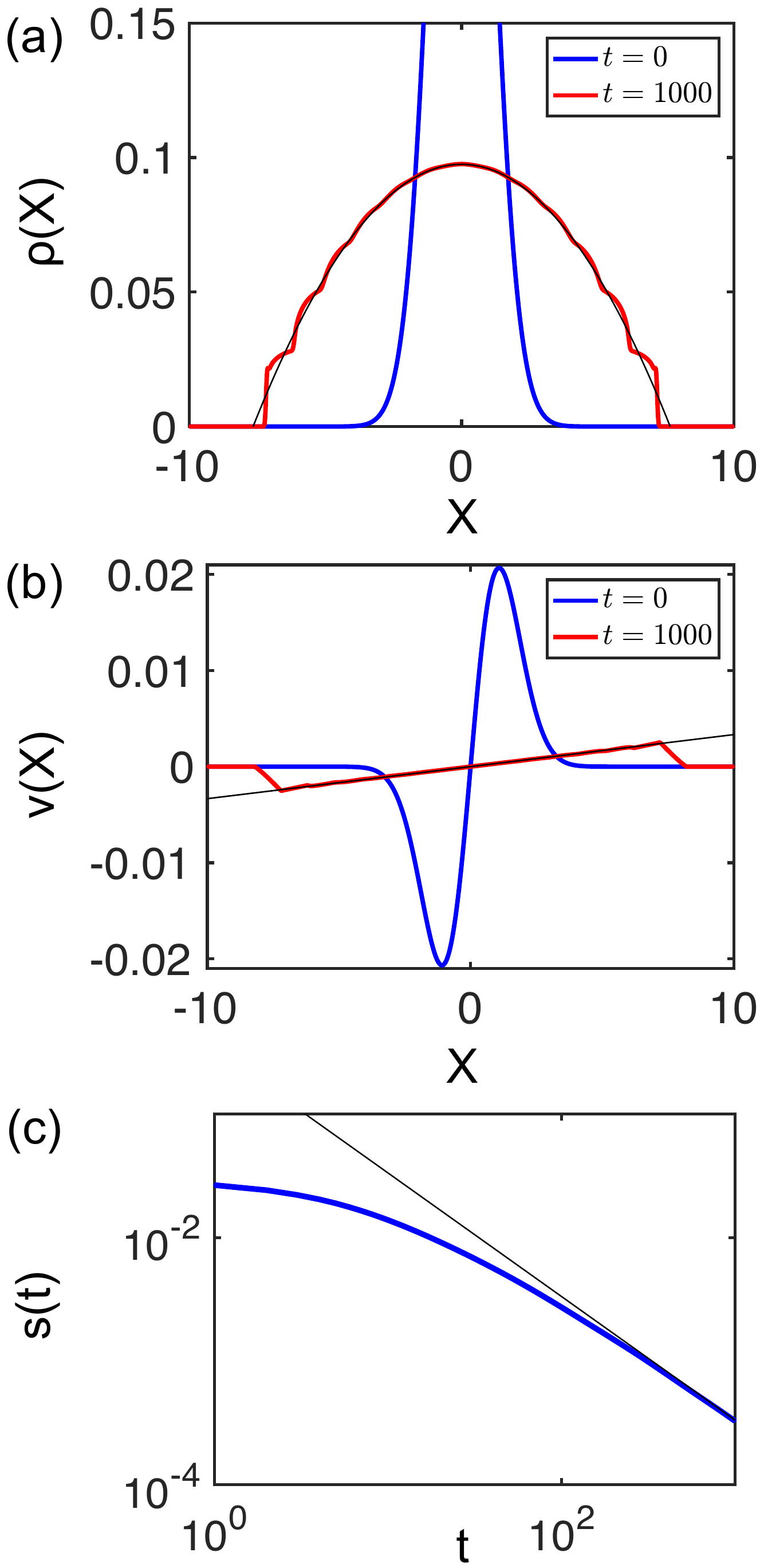}
\caption{Numerical resolution of Eq.\ref{eq2} and Eq.\ref{eq3}. (a) Density as function of space at time $t=0$ (blue) and $t=1000$ (red). The black line is a fit of equation $y(x)=-ax^2 + b$ with $a=0.00166$ and $b=0.0974$. (b) Velocity as function of space at time $t=0$ (bue) and $t=1000$ (red). The black line is a fit of equation $y(x)=cx$ with $c = 1/3000 $. (c) Slope $s(t)$ of the velocity as function of time, s(t) converges towards the line of equation $y(t) = 1/(3t)$ (black line).}
\label{figureA1}
\end{center}
\end{figure}

To find couples of solutions $(v(x,t), \rho(x,r))$ that satisfies both Eqs.\ref{eq2} and \ref{eq3}, we numerically solve those two equations  using as initial conditions: 
\begin{equation}
\begin{aligned}
\begin{gathered}
\rho(x,t=0) = \frac{e^{-\frac{x^2}{2}}}{\sqrt{2\pi}}\\ 
v(x,t=0) = v_p \int^{x+r_c}_{x-r_c}\sgn(x-x')\frac{e^{-\frac{x'^2}{2}}}{\sqrt{2\pi}}dx'
\end{gathered}
\end{aligned}
\end{equation}
In practice, we use a recursion method and calculate numerically
the density $\rho(x,t+dt)$  using Eq.\ref{eq2}, and then $v(x,t+dt)$ from $\rho(x,t+dt)$
using Eq. \ref{eq3}. As parameters we use: time step $dt=0.1$, space step $dx=0.01$, cutoff length $r_c=1$, particle velocity $v_p=0.1$ and total space length $L_M=100$.  
The numerical solutions $(v(x,t)$,$\rho(x,t))$ have the form: 
\begin{equation}
\begin{aligned}
\begin{gathered}
v(x,t) = \frac{cx}{t}\\
\rho(x,t) = -x^2 f(t) +g(t)
\end{gathered}
\end{aligned}
\end{equation}
with $c$ a constant and $f(t)$, $g(t)$ arbitrary functions of time,
and are shown in Fig.\ref{figureA1}.
Placing these solution in Eq.\ref{eq2} leads to:
\begin{equation}
\begin{aligned}
\begin{gathered}
f(t) = \frac{a}{t^{3c}}\\
g(t) = \frac{b}{t^c}
\end{gathered}
\end{aligned}
\end{equation}
with $a$ and $b$ constants. The density $\rho(x,t)$ then reads:
\begin{equation}
\rho(x,t) = -\frac{ax^2}{t^{3c}} + \frac{b}{t^c}
\end{equation}
We further use the above expression of $\rho$ in Eq.\ref{eq3}, which gives:
\begin{equation}
v(x,t) = 2av_p r_c^2 \frac{x}{t^{3c}}
\end{equation}
As a result, we obtain the equation $2 a v_p r_c^2 \frac{x}{t^{3c}} = \frac{cx}{t}$, which implies:
 \begin{equation}
\begin{aligned}
\begin{gathered}
c=1/3 \\
a = \frac{1}{6v_p r_c^2}
\end{gathered}
\end{aligned}
\end{equation}
To obtain the constant $b$ we impose the following constraints, namely the integral of the density over $L$ is equal to the total number of particles N, and the density is zero at $x=\pm L/2$,
\begin{equation}
\begin{aligned}
\begin{gathered}
N = \int^{L/2}_{-L/2} \rho(x,t)dx\\
\rho(\pm L/2,t) = 0
\end{gathered}
\end{aligned}
\end{equation}
which gives:
\begin{equation}
\begin{aligned}
\begin{gathered}
N = \frac{-aL^3}{12t}+ \frac{bL}{t^{1/3}}\\
\frac{-aL^2}{4t^2} + \frac{b}{t^{1/3}} = 0
\end{gathered}
\end{aligned}
\end{equation}
Finally, one has:
\begin{equation}
\begin{aligned}
\begin{gathered}
b =\bigg( \frac{3N^2}{32 r_c^2 v_p}\bigg)^{1/3}\\
L(t) = \bigg(36 N v_p r_c^2 t\bigg)^{1/3}
\end{gathered}
\end{aligned}
\end{equation}
and:
\begin{equation}
\begin{aligned}
\rho(x,t) &= \frac{-x^2}{6 v_p r_c^2 t} + \left(\frac{3N^2}{32 r_c v_p t}\right)^{1/3}\\
v(x,t) &= \frac{x}{3t}
\end{aligned}
\end{equation}

\section{Appendix C: Numerical simulation }

In integrating Eq.~\ref{numeq1}, we consider two cases for the velocity $\bm{v}$,
depending on a cut-off distance $r_c$. For $r_{ji}>r_c$, we assume that $\bm{v}=\bm{v}_{exp}/2$ where $\bm{v}_{exp}$ is an empirical function that describes the relative velocity between a pair of interacting particles measured experimentally at $c_p=0.05\%$~\cite{junot2023large}. Note that to have the velocity field $\bm{v}$, one has to divide $\bm{v}_{exp}$ by 2 since $\bm{v}_{exp}$ is the result of the contribution of two particles. Thus, $\bm{v}_{exp}$ is given by:
\begin{equation}\label{eqfit}
\begin{gathered}
\begin{aligned}
v_{exp}(r_{ji},\theta_{ji}) &=  (a r_{ji} + b)\exp(-r_{ij}^2/8) + c \\
a &= (\theta_{ji}/1.22)^{2/3}(3.12+4.4) -3.12 \\
b &= -(\theta_{ji}/1.22)^{2/3}(45+24) + 45 \\
c &= -(\theta_{ji}/1.22)^{2/3}(0.3+0.4) + 0.3 \, \, \, .
\end{aligned}
\end{gathered}
\end{equation}

Since the interaction between two particles is essentially radial, we neglect the azimuthal contribution of the velocity and $\bm{v}_{exp}=v_{exp}\bm{e_r}$. 
At the end of a sequence of approach ($\theta_{ji} > \theta_l$), two particles come in close contact (side-by-side). In this situation, we observe that the pair exhibit a three-dimensional leap-frog dynamics \cite{Massana-Cid2019}, sliding on each other and ending tip-to-tip before repealing. 
To reduce the complexity of the simulation scheme, we did not consider this transitory leap-frog state, but rather take it into account as a "scattering" event. In simulations, two particles at a closer distance than $r_c$ are place at a relative position $r_{ji}=r_c$ and $\theta_{ji} = 0$ in one time step:
\begin{equation}
\begin{gathered}
\begin{aligned}
v_x &= \frac{r_c-\Delta x}{2dt}\\
v_y &= \frac{-\Delta y}{2dt}
\end{aligned}
\end{gathered}
\end{equation}
with  $\Delta x = x_i - x_j$, $\Delta y = y_i - y_j$.
We perform numerical simulations of a system of $N$ point-like particles in a 2D square box of size $L$ with periodic boundary conditions. We consider a second cut-off distance $r_l=20 \, \rm{\mu m}$ beyond which the particles no longer interact,
 and $r_c = 0.2 \rm{\mu m}$.
The particles follow the above mentioned dynamics, and we solve the equation of motion with an Euler scheme with a time step $dt=10^{-2}$ s. 

\end{document}